# EdgeWorkflowReal: An Edge Computing based Workflow Execution Engine for Smart Systems

Xuejun Li[1], Ran Ding[1], Xiao Liu[2], Jia Xu[1], Yun Yang[3] and John Grundy[4]
[1]School of Computer Science and Technology, Anhui University, Hefei, China
[2]School of Information Technology, Deakin University, Geelong, Australia
[3]School of Software and Electrical Engineering, Swinburne University of Technology, Melbourne, Australia
[4]Faculty of Information Technology, Monash University, Melbourne, Australia
xjli@ahu.edu.cn; xiao.liu@deakin.edu.au; yyang@swin.edu.au; john.grundy@monash.edu

*Abstract*—Current cloud-based smart systems suffer from weaknesses such as high response latency, limited network bandwidth and the restricted computing power of smart end devices which seriously affect the system's QoS (Quality of Service). Recently, given its advantages of low latency, high bandwidth and location awareness, edge computing has become a promising solution for smart systems. However, the development of edge computing based smart systems is a very challenging job for software developers who do not have the skills for the creation of edge computing environments. The management of edge computing resources and computing tasks is also very challenging. Workflow technology has been widely used in smart systems to automate task and resource management, but there does not yet exist a real-world deployable edge computing based workflow execution engine. To fill this gap, we present EdgeWorkflowReal, an edge computing based workflow execution engine for smart systems. EdgeWorkflowReal supports: 1) automatic creation of a real edge computing environment according to user settings; 2) visualized modelling of edge workflow applications; and 3) automatic deployment, monitoring and performance evaluation of edge workflow applications in a smart system. Video: https://youtu.be/2e7Vm7rM5Zc.

## I. INTRODUCTION

With rapid development of cloud computing and IoT (Internet of Things) technology, cloud-based smart systems have been widely used in various scenarios, such as smart logistics, smart health, smart transportation, and so on [1-3]. However, with the increasing number of connected IoT devices and the growing amount of generated data, current cloud-based smart systems severely suffer from high response latency, limited network bandwidth and the restricted computing power of smart end devices, which will significantly impact the system's QoS (Quality of Service). These weaknesses are hindering the application of smart systems in many areas which require low latency, high reliability and energy efficiency [2-4].

In recent years, edge computing has attracted attention from both academia and industry due to its advantages such as low latency, high bandwidth and location awareness. Therefore, edge computing has become a promising solution for smart systems [2, 3]. Fig. 1 shows an example of a typical edge computing based smart UAV (Unmanned Aerial Vehicle) last-mile delivery system. This system can be divided into three layers: UAV layer, edge server layer and cloud server layer. The UAV layer is responsible for sensing and gathering the environment data during the flight mission. Due to the limited computing power and restricted battery life of the UAV, computation offloading strategies need to decide where the computing tasks should be executed based on the characteristics of computing tasks, for example, offloaded to the Cloud, offloaded to the Edge, or executed locally on the UAV itself. We can see the dependencies among computing tasks in the receiver recognition process which is represented as a DAG (Directed Acyclic Graph) based workflow application. The key issue to improve the QoS for edge computing based smart system is to effectively manage different computing resources and computing tasks. However, this can be a very challenging job for software developers who do not have the in-depth knowledge and skills for edge computing. Workflow technology has been widely used to automate the task and resource management. Therefore, an edge computing based workflow execution engine for smart systems is urgently needed.

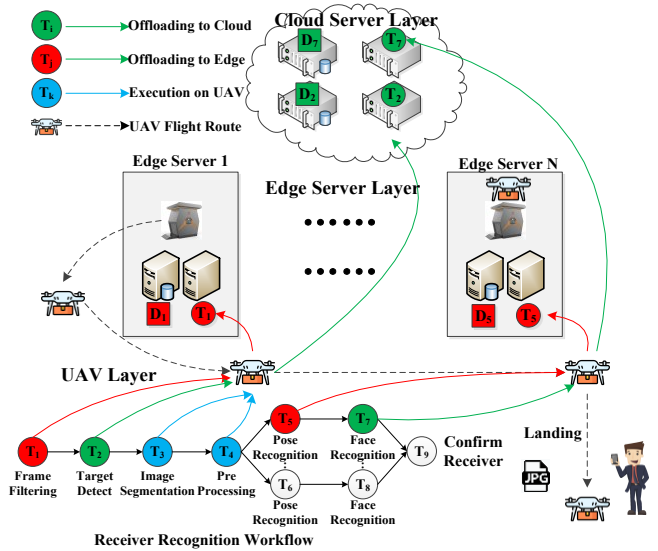

Fig. 1. An edge computing based smart UAV last-mile delivery system.

There are many research works devoted to developing workflow engines for business or scientific workflows (e.g. [1, 5, 6]). Unfortunately, existing workflow engines are mostly designed for cloud computing which are not suitable for edge computing because of the significantly different computing environment. So far, there are only a few simulation toolkits available such as iFogSim [7] and FogWorkflowSim [1] which are used for research purpose but cannot support the execution of workflow tasks in a real edge computing environment.

To fill this critical gap, in this paper, we propose EdgeWorkflowReal, which is an edge computing based workflow execution engine for smart systems. EdgeWorkflowReal supports: 1) automatic creation of a real edge computing environment according to user settings; 2) visualized modelling of edge workflow applications with various workflow structures and real computing tasks; 3) automatic deployment, monitoring and performance evaluation of edge workflow applications (e.g. task execution time and cost, and energy consumption of end devices) in user created edge computing environment. To demonstrate its capability, we provide a Web UI so that users can access EdgeWorkflowReal remotely and run their created edge workflow applications on our computing infrastructure.

## II. ARCHITECTURE OF EDGEWORKFLOWREAL

EdgeWorkflowReal enables the creation of a real edge computing environment and the execution of real workflow tasks. The simulation function for the evaluation of different resource and task management strategies is inherited from our previous work FogWorkflowSim [1]. The architecture of EdgeWorkflowReal is shown in Fig. 2. Specifically, it can be divided into three layers.

**The User Interface Layer** consists of three modules: workflow setting module, edge computing environment setting module, and strategy setting module. The major functions of the workflow setting module are: 1) visually creating workflow models with customised workflow structures; and 2) binding the workflow task model with real computing tasks. The function of the edge computing environment setting module is to configure the computing and network resources. The function of the strategy setting module is to select the task offloading strategy and task scheduling algorithm with different optimisation objectives such as time, energy and cost.

**The Workflow Management System Layer** is composed of the simulation module and the workflow engine. The simulation module is to facilitate the performance evaluation of different resource and task management strategies using a simulated edge computing environment so that users can select the most suitable strategies for their workflow applications before running them in the real edge computing environment which they need to pay for use of computing resources. Specifically, the simulation model is composed of three sub-modules, which are resource management module, performance metrics module, and the workflow visualization modeler module. The workflow engine module is composed of six sub-modules: workflow parsing module, scheduling algorithm module, workflow scheduler module, job execution module, container management module, and controller module. The workflow parsing module can analyze the standard workflow XML files and generate executable workflow models. The function of the scheduling algorithm module is to invoke the algorithms selected by the controller module according to the user settings. The workflow scheduler module is to manage the computation offloading and schedule workflow tasks on the computing resources according to the offloading and scheduling plans generated by the optimisation algorithms. The task execution module is responsible for workflow task execution in the edge computing environment according to the instructions from the workflow scheduler module. The design of the container management module refers to the computing resource management in the KNIX system [5]. All sub-modules in the workflow execution engine module are managed and coordinated by the controller module.

**The Database and Computing Resource Layer** is composed of MySQL database and container resource pool. The MySQL database stores various system data generated by EdgeWorkflowReal, such as the data for workflow structure, the data for edge computing resources, the data for workflow tasks, and so on. The container resource pool can generate various types of computing resources in the edge computing environment to support workflow task execution.

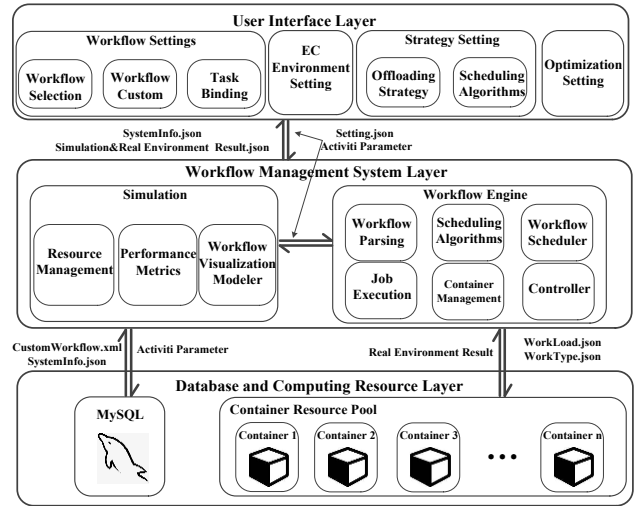

Fig. 2. Architecture of EdgeWorkflowReal.

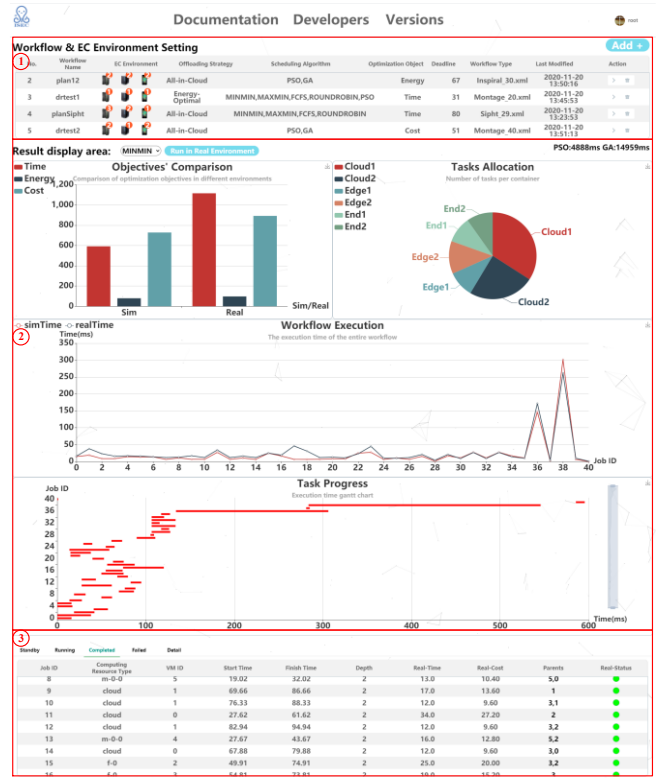

Fig. 3. Main page of EdgeWorkflowReal Web UI.

## III. IMPLEMENTATION

EdgeWorkflowReal is implemented in Java and the source code is available on Github[1]. To demonstrate its capability, we provide a Web UI [2] so that users can access EdgeWorkflowReal remotely and run their created edge workflow applications on our computing infrastructure. As shown in Fig. 3, the main page of EdgeWorkflowReal is divided into three areas: workflow and edge computing environment setting (Area 1), details for the execution result (Area 2) and the dynamic runtime monitoring (Area 3). To create a new edge workflow application, the users can click the "Add" button in Fig. 3, then the panel for adding a new edge workflow application is provided as shown in Fig. 4 which includes four steps as described below.

---

[1] https://github.com/ISEC-AHU/EdgeWorkflowReal

[2] http://47.98.222.243/EdgeWorkflowReal

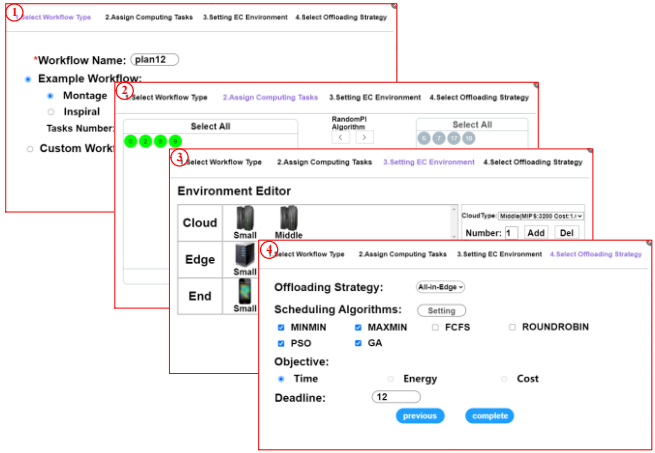

Fig. 4. Steps for adding a new edge workflow application.

### A. Select Workflow Type

As shown in Fig. 4, Step 1, the users need to create the workflow structure and workflow tasks. The workflow structure can be generated by using either provided example workflows or the visualized workflow modeler.

### B. Assign Computing Tasks

As shown in Fig. 4, Step 2, the workflow computing task models are to bind with real computing tasks. For demonstration purpose, four example scientific computing tasks are provided including: π value calculation, KMP character matching algorithm, Levenshtein string similarity detection algorithm, and selection sorting algorithm. The users can select the tasks from the unassigned task area and bind the task model with one of the real computing tasks. In the future, more built-in computing tasks will be included, and users will be able to upload their own computing tasks.

### C. Set Edge Computing Environment

As shown in Fig. 4, Step 3, the users need to set the computing resources in the edge computing environment for workflow execution. The computing resources are divided into three types: cloud server, edge server, and end device. The resources in each layer can be selected from three pre-configured types including Small, Medium, and Large, which represent different levels of computing power with different prices.

### D. Set Optimisation Algorithms

As shown in Fig. 4, Step 4, the users need to select the computation offloading strategy, the workflow scheduling algorithm, the optimisation objectives, and specify the deadline constraint. At the moment, the offloading strategy can be selected from the following three types: Energy-Optimal (energy optimal offloading strategy), All-in-Edge (all tasks are executed on edge servers), All-in-Cloud (all tasks are executed on cloud servers). The scheduling algorithms can be selected from MINMIN, MAXMIN, FCFS, ROUNDROBIN, PSO (Particle Swarm Optimisation), and GA (Genetic Algorithm) [1, 4]. Please note that the heuristic algorithms such as MINMIN, MAXMIN, FCFS and ROUNDROBIN are only used for time optimisation. The search algorithms such as PSO and GA can be used for multi-objective optimisation such as time, energy and cost.

### E. Generate and Evluate Execution plan using Simulator

After users complete all required settings, the edge workflow execution plan is generated and shown in Fig. 3, Area 1. Users can first start the simulation process of the plan to evaluate its performance. The evaluation results will support important decision-making such as whether the budget for current computing resources is enough to meet the deadline and which computation offloading strategy and task scheduling algorithm should be selected.

### F. Execute Edge Workflow in Real Environment

To run the edge workflow application in the real edge computing environment according to the execution plan, users can click the "Run in Real Environment" button to start the real workflow execution. As shown in Fig. 3, Area 3, the information of computing tasks in standby, running, completed, and failed status is displayed in the "Standby", "Running", "Completed" and "Failed" tabs respectively. Meanwhile, the information of all computing tasks can be accessed in the "Detail" tab. After all computing tasks are executed in the real edge computing environment, the four statistical graphs in Fig. 3, Area 2 will be produced. Specifically, the bar chart presents the comparison results under different objectives; the pie chart shows the percentage of tasks assigned to each computing resource; the line chart compares the execution time of all tasks (showing both simulation and real execution results); and the Gantt chart presents execution time span and sequence of all tasks.

## IV. EVALUATION

We describe the evaluation of EdgeWorkflowReal to demonstrate its effectiveness and usability. Firstly, the energy consumption of end devices, the task execution time and cost under various task scheduling algorithms are compared. Secondly, each workflow task's execution time, task assignment and execution Gantt chart are demonstrated.

### A. Experimental Environment and Parameter Settings

For demonstration purpose, EdgeWorkflowReal currently runs on Alibaba Cloud with the following configurations: Intel(R) Xeon(R) Platinum 8369HC dual-core 3.4GHz, 4G RAM, 40G ROM, and Ubuntu 18.04 64-bit operating system. The container uses Docker Version 19.03.6. The database is MySQL 5.7.31. EdgeWorkflowReal is developed in Java JDK 1.8. EdgeWorkflowReal can also be deployed in any user created computing environments for their smart systems as the computing resources will be managed by the KNIX system in the form of containers [5].

EdgeWorkflowReal supports various popular scientific workflow structures with different task numbers. Here, we use the Montage workflow [6] as an example, and compare various task scheduling algorithms under different optimisation objectives. The parameter settings of the edge computing environment are shown in Table I. Note that users can specify some of these parameters so that they can focus on specific objectives in the statistical graphs produced in Fig. 3. Currently, we have implemented six representative task scheduling algorithms as mentioned earlier. For the PSO algorithm, the particle number is 30. Both learning factors C1 and C2 are set as 2, and the inertia weight is set as 1. For the GA algorithm, the population size is 50. The rates of crossover and mutation are 0.8 and 0.1 respectively. The iteration numbers of PSO and GA are both 100. Energy-Optimal is the default computation offloading strategy.

TABLE I. PARAMETER SETTINGS OF EDGE COMPUTING ENVIRONMENT

| Parameters | End Device | Edge Server | Cloud Server |
|---|---|---|---|
| MIPS | 1000 | 1300 | 1600 |
| Running Power | 700 | 0 | 0 |
| Idle Power | 30 | 0 | 0 |
| Data Transmission Power (mW) | 100 | 0 | 0 |
| Data Receiving Power (mW) | 25 | 0 | 0 |
| Task Execution Cost ($) | 0 | 0.48 | 0.96 |
| Number of Device | 2 | 2 | 2 |

## B. Experimental Results and Analysis

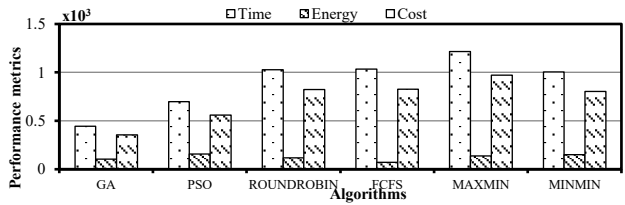

Fig. 5. Summary of experimental results.

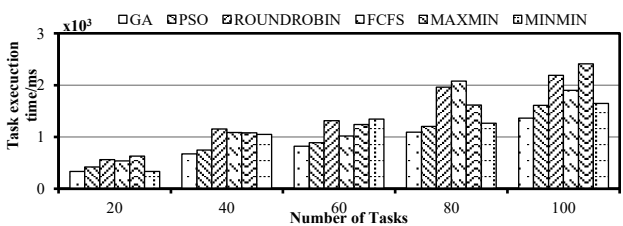

Fig. 6. Comparison of task execution Time.

The experimental results of six task scheduling algorithms on task execution time and cost, and energy consumption of the end devices are shown in Fig. 5. The detailed results of task execution time with different numbers of tasks are shown in Fig. 6. It can be clearly seen that the PSO and GA algorithms always achieve better performance on different objectives than other four algorithms. With the increasing number of tasks, the gap between GA and PSO is getting smaller. For example, when the number of tasks is 20, the task execution time under GA is 27% lower than PSO. When the number of tasks becomes 100, the gap is reduced to 18%. Therefore, GA is the most effective task scheduling algorithm for reducing task execution time under the current experimental settings.

## V. RELATED WORK

At present, there does not exist any research work on workflow execution engines in the edge computing environment. A number of simulators and toolkits was designed for researchers to investigate the resource management problem in the simulated edge computing (also known as fog) and cloud environment. Gupta et al. [7] proposed a fog environments simulator iFogSim. Liu et al. [1] designed and implemented an efficient fog computing simulation toolkit for automatically evaluating resource and task management strategies. Akkus et al. from Nokia Bell Labs [5] proposed SAND (now known as KNIX) which is an open-source, high-performance serverless platform for edge-enabled cloud applications. However, none of these works takes into account both the creation of real edge computing environment and the design of the workflow execution engine. In cloud computing environment, Calheiros et al. [8] proposed an extensible simulation toolkit named CloudSim. Balis [9] presented a HyperFlow scientific workflow engine which supports workflow models with complex structures. Altintas et al. [10] conducted the Kepler project which is an open-source scientific workflow management system. However, these cloud-based workflow engines are not suitable for the edge computing environment because of their significantly different computing and network architecture. To the best of our knowledge, EdgeWorkflowReal is the first workflow execution engine for real workflow tasks in the edge computing environment.

## VI. CONCLUSIONS AND FUTURE WORK

Edge computing is a promising solution for smart systems. An edge computing based workflow execution engine is essential for software developers to overcome the challenging issues such as the creation of edge computing environments, and the management of edge computing resources and computing tasks. In this paper, we have proposed EdgeWorkflowReal, which is the first edge computing based workflow execution engine for smart systems. Currently, EdgeWorkflowReal is being used for the development of a smart UAV delivery system as described in Fig. 1. In the future, we will expand its capability to support more built-in and user-submitted computing tasks so that it can be used for the development of many different edge computing based smart systems.

REFERENCES

[1] X. Liu, L. Fan, J. Xu, X. Li, L. Gong, J. Grundy and Y. Yang, "FogWorkflowSim: an automated simulation toolkit for workflow performance evaluation in fog computing," in *Proceedings of the 34th IEEE/ACM International Conference on Automated Software Engineering (ASE)*. IEEE, 2019, pp. 1114-1117.

[2] X. Wang, Z. Ning and S. Guo, "Multi-agent imitation learning for pervasive edge computing: a decentralized computation offloading algorithm," *IEEE Transactions on Parallel and Distributed Systems*, vol. 32, no. 2, pp. 411-425, 2020.

[3] X. Xu, Q. Huang, X. Yin, M. Abbasi, M. Khosravi and L. Qi, "Intelligent offloading for collaborative smart city services in edge computing," *IEEE Internet of Things Journal*, vol. 7, no. 9, pp. 7919-7927, 2020.

[4] J. Xu, X. Li, X. Liu, C. Zhang, L. Fan, L. Gong and J. Li, "Mobility-aware workflow offloading and scheduling strategy for mobile edge computing," in *Proceedings of the International Conference on Algorithms and Architectures for Parallel Processing*. Springer, pp. 184-199, 2019.

[5] I. Akkus, R. Chen, I. Rimac, M. Stein, K. Satzke, A. Beck, P. Aditya, et al., "SAND: towards high-performance serverless computing," in *Usenix Annual Technical Conference*, 2018, pp. 923-935.

[6] "WorkflowGenerator," https://confluence.pegasus.isi.edu/display/pegasus/WorkflowGenerator, accessed on 1st November, 2020.

[7] H. Gupta, A. Dastjerdi, S. Ghosh and R. Buyya, "iFogSim: a toolkit for modeling and simulation of resource management techniques in the Internet of Things, Edge and Fog computing environments," *Software: Practice and Experience*, vol. 47, no. 9, pp. 1275-1296, 2017.

[8] R. Calheiros, R. Ranjan, A. Beloglazov, C. Rose and R. Buyya, "CloudSim: a toolkit for modeling and simulation of cloud computing environments and evaluation of resource provisioning algorithms," *Software: Practice and Experience*, vol. 41, no. 1, pp. 23-50, 2011.

[9] M. Malawski, A. Gajek, A. Zima, B. Balis and K. Figiela, "Serverless execution of scientific workflows: experiments with hyperflow, AWS Lambda and Google cloud functions," *Future Generation Computer Systems*, vol. 110, pp. 502-514, 2020.

[10] J. Wang, M. AbdelBaky, J. Diaz-Montes, S. Purawat, M. Parashar, I. Altintas, "Kepler+ CometCloud: dynamic scientific workflow execution on federated cloud resources," *Procedia Computer Science*, vol. 80, pp. 700-711, 2016.